\documentclass[a4paper,11pt]{article}
\pdfoutput=1

\usepackage{jcappub}
\usepackage{physics}

\newcommand{\Si}{\operatorname{Si}}
\newcommand{\Ci}{\operatorname{Ci}}

\arxivnumber{2607.26174}

\title{A Periodically Interacting Dark Sector: Signatures and Constraints from CMB and Cosmic Expansion Data}

\author[a]{Marco Antonio Cardoso Alvarez,\note{Corresponding author.}}
\author[b,c]{Micol Benetti,}
\author[a]{Leila Graef,}
\author[d,e]{and Robert Brandenberger}

\affiliation[a]{Instituto de Física, Universidade Federal Fluminense,\\
24210-346 Niteroi, RJ, Brazil}
\affiliation[b]{Scuola Superiore Meridionale,\\
Via Mezzocannone 4, 80134 Naples, Italy}
\affiliation[c]{INFN Sezione di Napoli, Complesso Universitario di Monte S. Angelo,\\
Edificio G, Via Cinthia, 80126 Naples, Italy}
\affiliation[d]{Physics Department, McGill University,\\
Montreal, QC H3A 2T8, Canada}
\affiliation[e]{Trottier Space Institute, McGill University,\\
Montreal, QC H3A 2T9, Canada}

\emailAdd{marcocardoso@id.uff.br}

\abstract{We introduce a novel cosmological model that provides an effective description of the back-reaction of super-Hubble fluctuations on the cosmological background, a generic effect expected to arise in any cosmological scenario without requiring additional ingredients. At the phenomenological level, it can be interpreted as an interacting dark sector scenario in which the sign of the energy transfer changes periodically over time. We constrain the model using CMB, BAO, and Type Ia supernova data. We find that a significant amplitude of the oscillatory interaction is allowed by the data, although there is no statistically significant preference for this model over $\Lambda$CDM.}

\begin{document}
\maketitle
\flushbottom

\section{Introduction}%
\label{sec1}

The discovery of the accelerated expansion of the Universe \cite{Riess:1998cb,Perlmutter:1998np}, later confirmed with increasing precision by observations of the cosmic microwave background (CMB), baryon acoustic oscillations (BAO) and type Ia supernovae, established the $\Lambda$CDM model as the standard paradigm of modern cosmology. In this framework, the present epoch of accelerated expansion is attributed to a cosmological constant $\Lambda$, whose energy density is observed to be comparable to that of matter today. This apparent coincidence, together with the enormous mismatch between the observed value of $\Lambda$ and the value naively expected from quantum field theory, constitutes the cosmological constant problem \cite{Weinberg:1988cp} and has motivated a large body of work exploring alternatives to a pure cosmological constant, ranging from dynamical scalar field models of dark energy to modifications of gravity on cosmological scales (see e.g. \cite{Copeland:2006wr,Peebles:2002gy} for reviews).

A particularly well studied class of alternatives consists of models in which dark energy is not a separately conserved fluid, but instead exchanges energy with the dark matter sector \cite{Wetterich:1994bg,Amendola:2007yx,Li:2025owk,Li:2026xaz,Chakraborty_2025}. Beyond providing a phenomenological framework in which to test deviations from $\Lambda$CDM, such interacting dark sector models can help alleviate the coincidence problem, since the ratio of the dark energy and dark matter densities can be driven towards an attractor or scaling behaviour rather than depending sensitively on initial conditions. They have also been invoked in connection with the $H_0$ tension and other observational anomalies \cite{Wang:2016lxa,Bolotin:2013jpa,Valiviita:2008iv,DiValentino:2019ffd,DiValentino:2025sru,vonMarttens:2018iav,Yang:2018euj,Nunes:2022bhn, Benetti:2024dob, Benetti:2021div,Salzano:2021zxk, Benetti:2019lxu,Li:2026asg}. Importantly, an interaction between the dark sectors need not be introduced by hand: as we discuss below, it can also arise as an effective, derived consequence of physics that is otherwise well motivated -- in particular, of the back-reaction of cosmological perturbations on the background evolution of the Universe.

In this paper we study the background cosmology, the growth of structure and CMB anisotropies in an interacting dark sector model in which the sign of the energy transfer changes periodically in time on a Hubble time scale over the entire time interval between recombination and the present time.

There are several motivations for this specific study, which we now discuss in turn. Our initial motivation came from the back-reaction of cosmological perturbations on the background geometry. As shown in \cite{Mukhanov:1996ak,Abramo:1997hu},  each super-Hubble scale Fourier mode of the scalar metric fluctuations acts as a negative contribution to the effective cosmological constant.  This effect is physically measurable if the clock field with respect to which we measure the cosmological evolution is not comoving with the dominant fluid (see e.g. \cite{Geshnizjani:2003cn, Brandenberger:2018fdd}, and \cite{Brandenberger:2002sk} for a review of the early literature). This is the case in the usual late time cosmology where we measure time in terms of the temperature of the CMB, and the dominant matter field is not constant on the fixed temperature hypersurfaces. As shown in \cite{Abramo:1997uy},  in this case the continuity equation of matter and of the effective cosmological constant are not separately conserved, and there is an effective energy flow between the dark energy and the dark matter components.  As speculated in \cite{Brandenberger:2002sk},  the interplay between back-reaction terms and background cosmology may lead to a ``scaling solution'' in which the effective cosmological constant oscillates about its average value on a Hubble time scale \footnote{The idea that oscillating dark energy might provide insights on the cosmological constant and the coincidence  problems was mentioned in \cite{Rubano:2003er, Linder:2005dw, Nojiri:2006ww}.}.

Our setup leads to a cosmology in which the equation of state of the joint matter - dark energy fluid has an oscillatory behaviour. The idea that the dark energy equation of state might be oscillatory is old (see e.g. \cite{Rubano:2003er, Linder:2005dw}). There have been studies aiming at reconstructing the late time dark energy equation of state parameter from background cosmology observations which yield tentative evidence for oscillations on a Hubble time scale (see e.g. \cite{Zhao:2017cud,Zhang:2019jsu,Escamilla:2024fzq}, and \cite{Colgain:2021pmf} for caveats on the interpretation), and there have been studies providing observational constraints on possible oscillations \cite{Kurek:2007bu,Jain:2007fa,Pan:2017zoh,Rezaei:2019roe,Rezaei:2024vtg}.  Oscillating dark energy scalar fields might yield an oscillating equation of state (see e.g. \cite{Zhao:2005vj,Lazkoz:2007mx}) \footnote{After the first draft of this article was written, a paper appeared \cite{Hussain:2026srf} which studies a phenomenological model in which the equation of state of dark energy oscillates on a Hubble time scale with an amplitude which decreases in time.}.  Another setup which yields dark energy with an oscillating equation of state is the \textit{everpresent Lambda} scenario \cite{Ahmed:2002mj} motivated by the causal set approach to quantum gravity (see \cite{Bombelli:1987aa} for an original article, and \cite{Dowker:2003hb,Surya:2019ndm} for reviews). In this case, $\Omega_{\Lambda}$, the fractional contribution of the dark energy to the total energy, oscillates about the value zero.  In this setup, a best fit parameter estimation was performed in \cite{Zwane:2017xbg}.

What is special in our scenario is that the oscillations stem from an oscillatory interaction between the effective dark energy and the matter sector (see \cite{Saez-Gomez:2008mkj} for related ideas), and that there is a good reason that the time scale of oscillations is the Hubble time scale (see appendix~\ref{App:A}). There has been a lot of work (see e.g. \cite{Buen-Abad:2017gxg,Giare:2024ytc,Wang:2024vmw,Borges:2023xwx,Johnson:2021wou,Benetti:2021div,Yang:2020uga,Johnson:2020gzn,Benetti:2019lxu,Wang2016,Ferreira:2014jhn,Chimento:2013rya,Costa:2013sva,Abdalla:2012ug,Costa:2012xf,Pavan:2011xn,Clemson2012,He:2010im,Baldi:2010vv,Baldi:2008ay,Feng:2008fx,Wang:2007ak,Guo:2007zk,Wang:2006qw,Amendola:2007yx,Barnes:2005bn,Das:2005yj,Wang:2005ph,Wang:2005jx,Huey:2004qv,Chimento:2003iea,Wetterich:1994bg,Li:2025owk,Li:2026xaz,Li:2026asg}) on phenomenological interacting dark sector models in which there is a continuous energy transfer between the dark matter and dark energy sectors given by a transfer rate of the form
\begin{equation}
Q \, = \, \zeta {\cal{H}} \rho_{\rm x} \, ,
\end{equation}
where ${\cal{H}}$ is the Hubble expansion rate in conformal time, $\rho_x$ is the energy density in one of the fluids, and $\zeta$ is a constant. For example, in the ``dark dimension'' scenario \cite{Bedroya:2025fwh,Anchordoqui:2022svl,Gonzalo:2022jac,Montero:2022prj,Agrawal:2019dlm} (motivated by superstring theory), the time dependence of the radion (the radius of an extra dimension) leads to a time-dependent dark matter mass. If the radion is oscillating about its ground state, this will induce a periodically varying mass, and hence a periodically oscillating value of $\zeta$.  See also \cite{Chen:2025ywv,Khoury:2025txd,CarrilloGonzalez:2017cll, Khoury:2026svx} for similar approaches, and \cite{Pereira:2026llu,  Jiang:2026cqh,  Teixeira:2026yjd, Jensko:2026taf, Zhai:2026uwr} for some very recent work. A further distinguishing feature of our scenario, relative to the models discussed above, is that the interaction is oscillatory as a function of redshift, rather than sustained and monotonic.

Since the evolution of the cosmology between the time of recombination and the present time deviates from what is proposed in the vanilla $\Lambda$CDM scenario, we find a non-trivial contribution to the Integrated Sachs-Wolfe (ISW) effect (see e.g. \cite{Hu_2002,Aghanim_2008,Nishizawa_2014} for reviews) as would be expected (see figure~\ref{fig3}).

We will work in the context of a spatially flat Friedmann-Lemaitre-Robertson-Walker model. The cosmological scale factor is written as $a(t)$, where $t$ is the physical time, and we will mostly be using the cosmological redshift $z(t)$ instead of time $t$. The Hubble expansion rate is denoted by $H(z)$. The energy density of a cosmological component $X$ is denoted by $\rho_X(z)$ and its pressure by $p_X(z)$.  As usual, the equation of state parameter of the fluid $X$ is given by $w_X(z) \equiv p_X(z) / \rho_X(z)$. Following the usual notation, the contribution of the fluid $X$ to the total current energy density is given by $\Omega_X \equiv \rho_X(z = 0) / \rho_c(z = 0)$, where $\rho_c(z = 0)$ is today's total energy density under the assumption that the spatial curvature vanishes.

Having situated our work within this broader context, from this point onward we focus specifically on the scenario motivated by the back-reaction of super-Hubble fluctuations.

\section{Model}
\label{sec:Model}

We build the background cosmology of our model by including the standard components, i.e. radiation, a cosmological constant, and pressureless matter (both baryonic and dark), together with a back-reaction term that interacts with the matter sector and is characterized by the equation of state $p = -\rho$. Such a term might arise from the back-reaction of long-wavelength cosmological perturbations. We denote the corresponding energy densities by $\rho_r$, $\rho_{\Lambda}$, $\rho_m$ and $\rho_{BR}$. We can combine the last two terms as
\begin{equation}\label{eq:rho_M}
\rho_{\rm M}(z) \equiv \rho_{\rm m}(z) + \rho_{\rm BR}(z) \, 
\end{equation}
where
\begin{equation}
    \rho_{\rm BR}(z) \equiv A_{\rm BR}\rho_{\rm m0}\cos{(f_{\rm BR}z)}(1+z)^3
\end{equation}
and write the background evolution as
\begin{equation}
H(z) = H_0 \sqrt{\Omega_{\rm r0}(1+z)^4 + \Omega_{\rm \Lambda} + \Omega_{\rm m0}\, y(z) + \Omega_{\rm BR}(z)} \, , \label{eq:H_ref_form_m_case}
\end{equation}
where the subscript ``0'' denotes quantities evaluated at $z = 0$. The last two terms on the right-hand side are the sum of the contributions of matter and the back-reaction terms respectively. In the absence of energy exchange between the two components, the energy density in $\rho_M$ would scale as $(1+z)^3$, just like usual matter; in the presence of energy exchange, however, the matter density scaling is more complicated and is encoded in the function $y(z)$. We refer to this model as the BR, back-reaction, model.

Our key assumption is that the back-reaction contribution, $\rho_{\rm BR}$, tracks the matter contribution, but oscillates about it on a time scale set by the Hubble expansion time, i.e. it oscillates as $\cos(f_{\rm BR}z)$, where $f_{\rm BR}$ is a constant of order $1$ \footnote{This time dependence is motivated by the dynamics of the interplay between the effective energy density from super-Hubble cosmological perturbation modes and the matter sector \cite{Brandenberger:2002sk} --- see appendix~\ref{App:A} for a review.}. We define the back-reaction density as
\begin{equation}
\label{eq:rhoBR_def_ref}
\rho_{\rm BR}(z)  =\rho_{m0} A\cos(f_{\rm BR}z)\,(1+z)^3.
\end{equation}
Here $A$ is a dimensionless constant giving the relative amplitude of the back-reaction term with respect to matter. For $A \ll 1$, the back-reaction energy density tracks the matter energy density to leading order in $A$ \footnote{Note that the continuity equation is not satisfied by the dark energy fluid individually, since there is continuous energy exchange between dark energy and dark matter. The equation of state of the dark energy fluid is $w = -1$ at all times, while $\rho_{\rm BR}$, averaged over time, decays as matter. Because of its equation of state, it is justified to regard this term as an effective (time-dependent) cosmological constant.}.

The continuity equation for the sum of the matter and the back-reaction contributions requires
\begin{equation}
\dot{\rho}_{\rm M}(t)+3H(t)(1+w_{\rm M}(z))\rho_{\rm M}(t)=0, \label{eq:effective_matter_continuity_equation}
\end{equation}
where $w$ is the equation of state of the sum of the two fluids. Based on the results for the contribution of super-Hubble fluctuation modes to the effective equation of state \cite{Mukhanov:1996ak,Abramo:1997hu,Brandenberger:2002sk} we have
\begin{equation}
p_{\rm M} \, = p_{\rm m} + p_{\rm BR} = - \rho_{\rm BR}
\end{equation}
and hence
\begin{equation}
w_{\rm M}(z)=-\frac{\rho_{\rm BR}(z)}{\rho_{\rm M}(z)}. \label{eq:back_reaction_continuity_equation}
\end{equation}

\begin{figure}[htbp]
\centering
\includegraphics[width=0.8\textwidth]{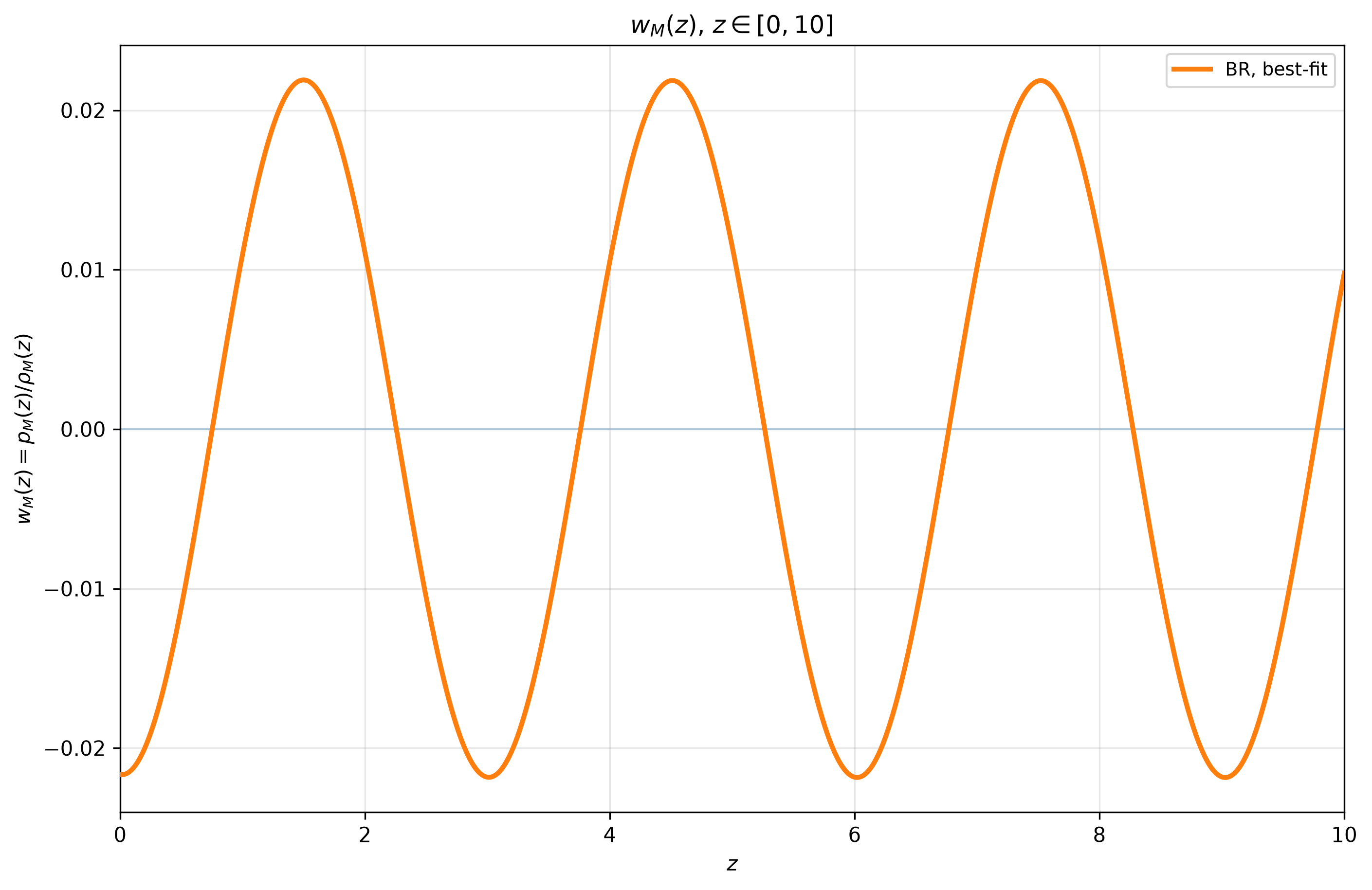}
\par\medskip
\includegraphics[width=0.8\textwidth]{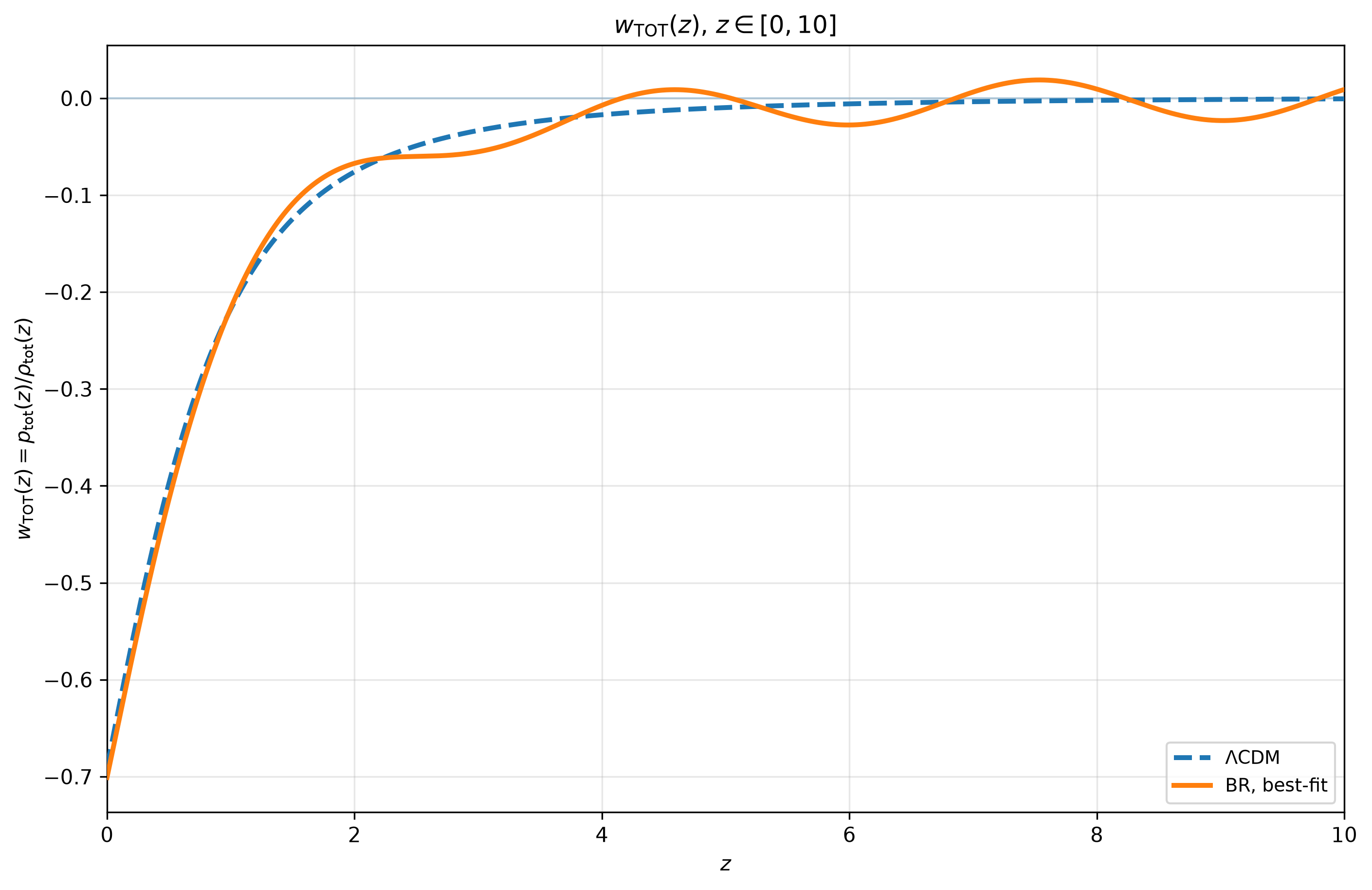}
\caption{Top: Equation of state of the matter fluid, $w_M(z)$ as a function of redshift. Bottom: Equation of state for the totality of matter, $w_{\rm TOT}$, as a function of redshift. For both figures, the best-fit values used were $A_{\rm BR} = 0.02$ and $f_{\rm BR} = 2$.}
\label{fig1}
\end{figure}

We plot the redshift dependence of the matter equation of state $w_M(z)$ and of the total equation of state $w_{\rm TOT}$ (which includes the contribution from the vacuum energy $\Lambda$ and radiation terms) in figure~\ref{fig1}. The matter equation of state oscillates about the value $w = 0$ for pressureless dust, while at very low redshifts the total equation of state tends to that of a cosmological constant.

If we compare our scenario to the usual interacting dark sector models \cite{Amendola2000,Zimdahl2001,Clemson2012,Wang2016}, in our case the mixing term $Q$, defined by
\begin{equation}
Q \, = \dot{\rho}_{\rm BR} + 3H(1+w_{\rm BR})\rho_{\rm BR} = \dot{\rho}_{\rm BR} \, ,
\end{equation}
is given by
\begin{equation}
Q \, = \, A_{\rm BR} H \rho_{\rm m0} (1+z)^3\bigl[ f_{\rm BR}(1 + z) \sin(f_{\rm BR}z) - 3 \cos(f_{\rm BR}z)\bigr] \, .
\end{equation}

We now derive the exact background evolution $H(z)$ of the model. Combining eq.~\eqref{eq:effective_matter_continuity_equation} with eq.~\eqref{eq:back_reaction_continuity_equation} gives
\begin{equation}
\dot{\rho}_{\rm M}+3H\rho_{\rm m} \, = \, 0, \label{eq:contM_step3}
\end{equation}
which is equivalent to \footnote{using
\begin{equation}\label{eq:dt_dz_relation}
\dv{}{t} \, = \, -(1+z)H(z)\dv{}{z}.
\end{equation}}
\begin{equation}\label{eq:cont_split_z}
\dv{\rho_{\rm m}}{z}+\dv{\rho_{\rm BR}}{z}-\frac{3}{1+z}\rho_{\rm m}=0.
\end{equation}

\begin{figure}[htbp]
\centering
\includegraphics[width=0.8\textwidth]{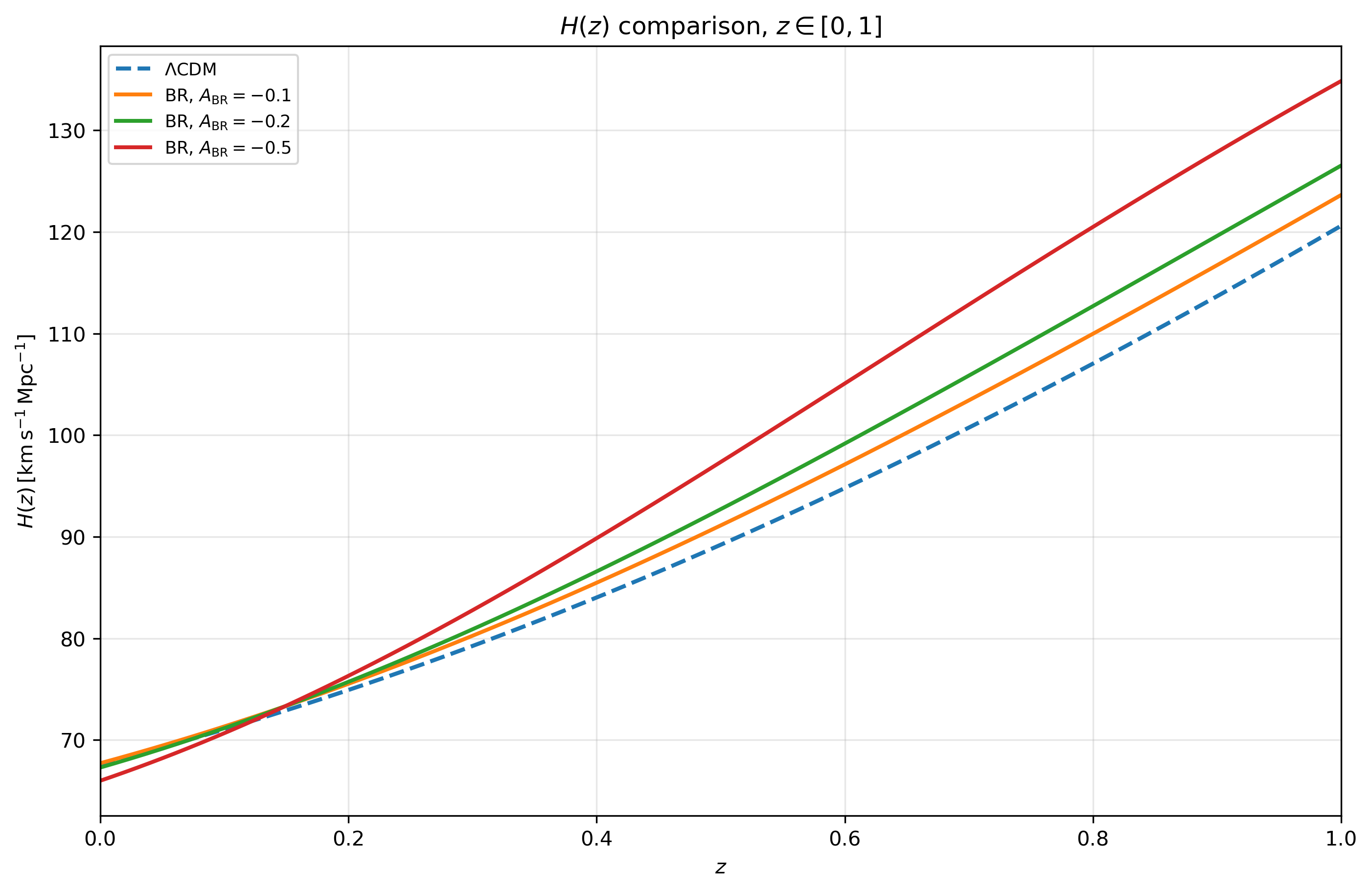}
\par\medskip
\includegraphics[width=0.8\textwidth]{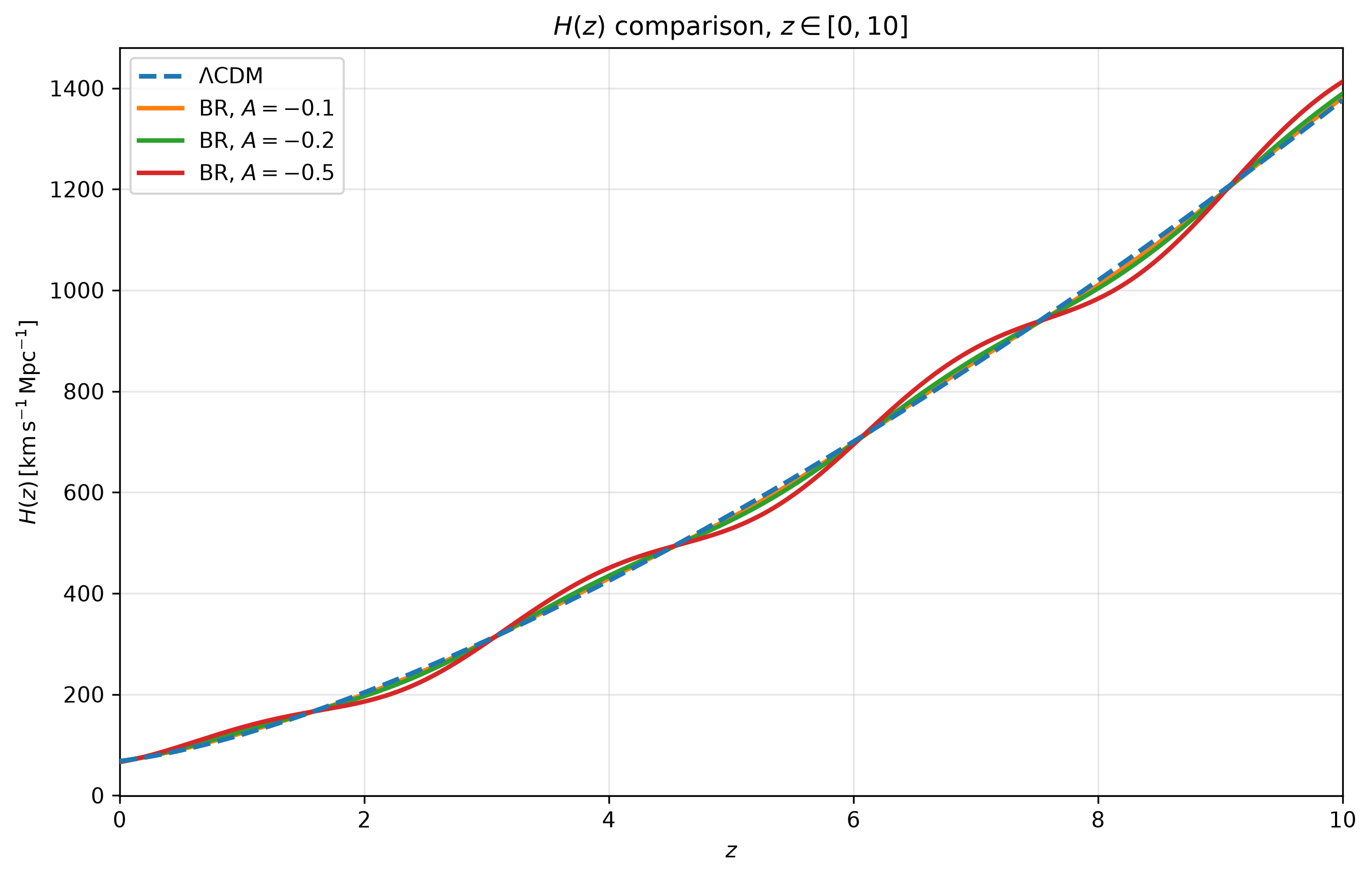}
\caption{Comparison of the Hubble expansion rates for the $\Lambda$CDM model and some BR models, fixing $f_{\rm BR}=2$ and using $A_{\rm BR}=-0.1$, $-0.2$, and $-0.5$. Top: we consider the range $0\leq z\leq 1$. Bottom: we consider the range $0\leq z\leq 10$.}
\label{Fig2}
\end{figure}

Using the definition above, we obtain
\begin{equation}\label{eq:rhotildem_ODE_dimful_m_case}
\begin{split}
    \frac{d\rho_{\rm m}}{dz}-\frac{3}{1+z}\rho_{\rm m}
    \, = \, A_{\rm BR}\rho_{\rm m0}\Big[f_{\rm BR}(1+z)^3\sin(f_{\rm BR}z)
    -3(1+z)^2\cos(f_{\rm BR}z)\Big] .
\end{split}
\end{equation}
This yields the differential equation for the dimensionless growth function $y(z)$ of eq.~\eqref{eq:H_ref_form_m_case}:

\begin{equation}\label{eq:y_ODE_m_case}
\begin{split}
    \frac{dy}{dz}(z)-\frac{3}{1+z}\,y(z)\,=\,-A_{\rm BR}(1+z)^2\Big[3\cos(f_{\rm BR}z)
    - f_{\rm BR}(1+z)\sin(f_{\rm BR}z)\Big],
\end{split}
\end{equation}
whose exact solution is
\begin{align}
\label{eq:y_exact}
y(z) &= (1+z)^3\Big\{1-A_{\rm BR}\cos(f_{\rm BR}z)\\
&-3A_{\rm BR}\Big[\cos(f_{\rm BR})\left(\Ci(x)-\Ci(x_{\rm rec})\right)+\sin(f_{\rm BR})\left(\Si(x)-\Si(x_{\rm rec})\right)\Big]\Big\}.
\end{align}
where
\begin{equation}
x \, \equiv \, f_{\rm BR}(1+z).
\end{equation}
The calculation leading to the solution above is given in appendix~\ref{app:B}.

In figure~\ref{Fig2} we illustrate the resulting evolution of $H(z)$, compared with the standard $\Lambda$CDM model. In the bottom pannel we note that $H(z)$ oscillates about the value for the reference cosmology, with the oscillation time scale set by the Hubble time and the constant $f_{\rm BR}$. We see that the maximal deviation of $H(z)$ is set by the amplitude $A_{\rm BR}$. 

\section{Method}
\label{sec:Method}

In order to constrain our model with cosmological data, we implement it in the Cosmic Linear Anisotropy Solving System (CLASS) code~\cite{Blas:2011rf} according to the discussion in section~\ref{sec:Model}. We use the usual baryon and CDM background densities together with an additional effective background component, as described in the previous section and explicitly calculated in appendix~\ref{app:B}. This is a new effective matter-like correction on top of the standard matter budget. In addition, the implementation assigns to this extra component its corresponding effective pressure, together with the associated logarithmic derivative entering the total pressure evolution. Note that, at this first step, we are only interested in modifying the standard cosmology at the background level, but that doesn't mean that the perturbative sector won't be affected. When total quantities (summed through all components) are considered, our model will depart from $\Lambda$CDM even at the level of perturbative calculations. Finally, this correction is switched off for $z\ge z_{\rm rec}$, so that the model matches the standard matter behavior prior to recombination and departs from it only in the late-time background evolution.

\begin{figure}[htbp]
\centering
\includegraphics[width=0.8\textwidth]{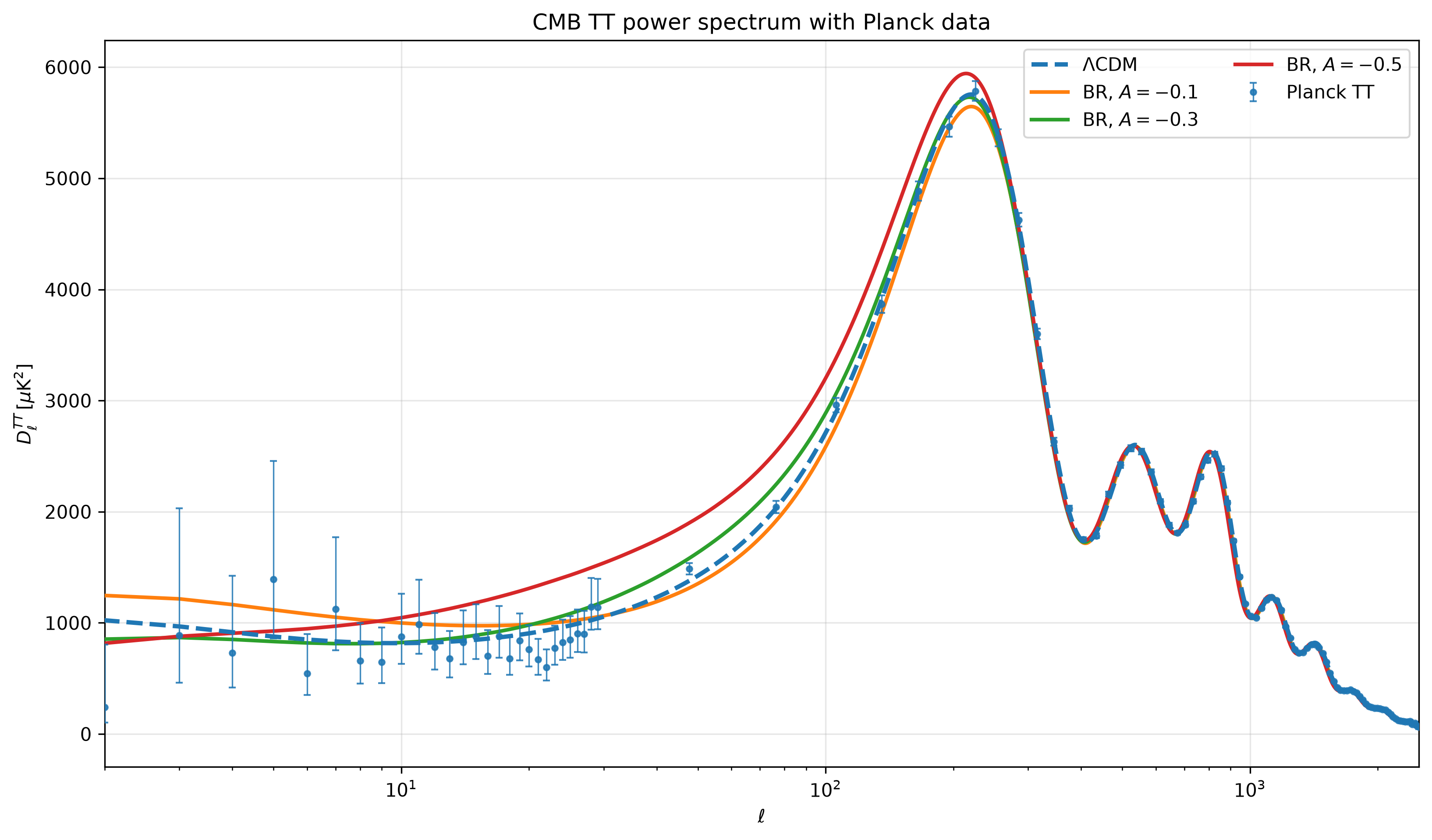}
\par\medskip
\includegraphics[width=0.8\textwidth]{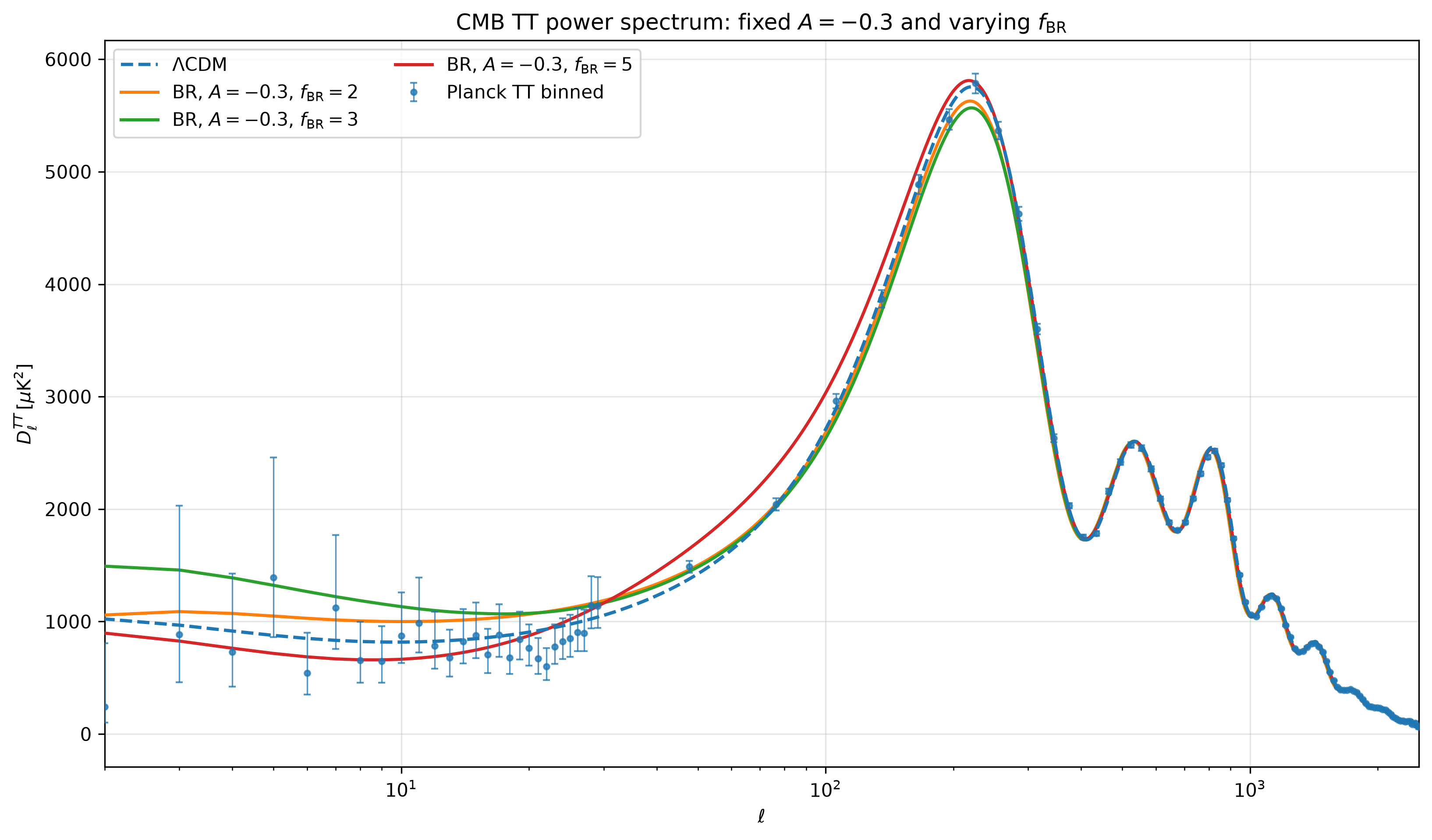}
\caption{Angular power spectrum of CMB temperature anisotropies (TT spectrum) in our scenario, compared to the predictions in the reference $\Lambda$CDM cosmology. Top: we fix $f_{\rm BR}=2$ and vary $A_{\rm BR}$. Bottom: we vary $f_{\rm BR}$ for an exaggerated fixed value of $A_{\rm BR}$.}
\label{fig3}
\end{figure}

The predictions for the temperature anisotropy power spectrum are shown in figure~\ref{fig3}: in the top pannel, the amplitude $A_{\rm BR}$ of the coupling is varied for a fixed value of the frequency $f_{\rm BR}$; in the bottom one, $f_{\rm BR}$ is varied for fixed $A_{\rm BR}$. Data points from the Planck survey \cite{Planck:2018vyg} are also shown. As can be seen, the model affects the low multipoles and produces a contribution to the ISW effect. This is expected since the evolution of $H(z)$ in our scenario departs from that of the $\Lambda$CDM model over the entire interval between recombination and the present, and this deviation translates into a variation of the gravitational potential traversed by CMB fluctuations along the line of sight. Nevertheless, the contributions partially cancel because $H(z)$ oscillates about the reference $\Lambda$CDM value, rather than deviating with the same sign throughout the evolution.

The ISW contribution predicted in our scenario depends on the parameters $A_{\rm BR}$ and $f_{\rm BR}$. Changing the value of $A_{\rm BR}$ simply changes the amplitude of the deviation in the CMB temperature angular power spectrum relative to the $\Lambda$CDM model. The dependence on $f_{\rm BR}$ (see bottom panel of figure~\ref{fig3}) leads to a change in the angular scale of the minimum amplitude of the power spectrum.

For our analysis, we consider Cosmic Microwave Background (CMB) data from the Planck 2018 release, specifically the Plik ``TT, TE, EE + lowE'' likelihood, which combines the temperature power spectra and TE and EE cross-correlations over the range $\ell \in [30, 2508]$, the low-$\ell$ temperature likelihood, and the low-$\ell$ SimAll EE likelihood, as well as its lensing reconstruction power spectrum \cite{Planck:2019nip, Planck:2018lbu}. We also include supernova data from the Pantheon+ survey \cite{Scolnic:2021smi, Brout:2022vxf}, consisting of 1701 light curves in the redshift range $0.001 < z < 2.26$, which constrain the late-time background cosmology. Following standard practice, we impose a Gaussian prior on the absolute magnitude $M_\text{B}$ as calibrated by local distance ladder measurements \cite{Riess:2020fzl}. Additionally, we incorporate Baryon Acoustic Oscillation (BAO) measurements from the 6dFGS \cite{Beutler:2011hx, Beutler:2012px}, SDSS DR7 MGS \cite{Ross:2014qpa}, and BOSS DR12 samples \cite{BOSS:2016wmc}. When we refer to the ``full'' dataset, we mean the joint cosmological data cited above, while we refer to ``CMB+BAO'' when using this reduced set of data of Planck 2018 and BAO.

We perform a Monte Carlo Markov Chain (MCMC) analysis \cite{MCMC,Lewis:2002ah, Lewis:2013hha} using the Cobaya code \cite{Torrado:2020dgo}. 
Our theoretical model includes the standard cosmological parameters: the physical baryon density, $\omega_{\text{b}}=\Omega_{\text{b}}h^2$, the physical cold dark matter density, $\omega_{\text{cdm}}=\Omega_{\text{cdm}}h^2$, the optical depth, $\tau_{\text{reio}}$, the primordial scalar amplitude, $A_{\text{s}}$, the primordial spectral index, $n_{\text{s}}$, and the Hubble constant, $H_0$, along with the back-reaction parameters $A_{\rm BR}$ and $f_{\rm BR}$. The standard cosmological parameters are allowed to vary with the usual priors, while for the two  additional free parameters we adopt the priors $A_{\rm BR}\in[-0.2,0.2]$ and $f_{\rm BR}\in[1,5]$.

\begin{figure}[htbp]
\centering
\includegraphics[width=0.9\textwidth]{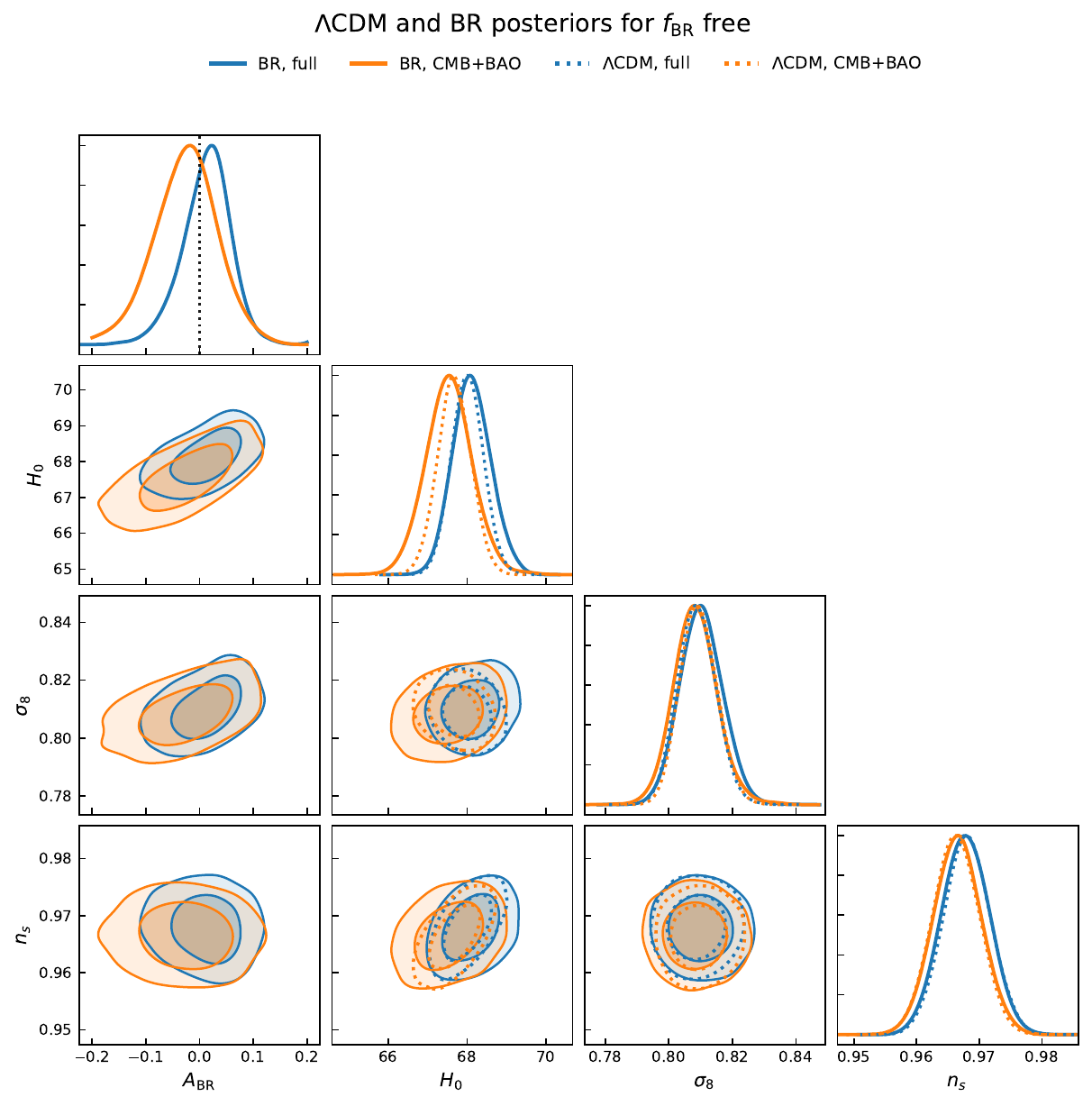}
\caption{Triangle plot showing the constraints on the model and cosmological parameters in both our scenario and in the reference $\Lambda$CDM model.}
\label{fig4}
\end{figure}

\begin{figure}[htbp]
\centering
\includegraphics[width=0.8\textwidth]{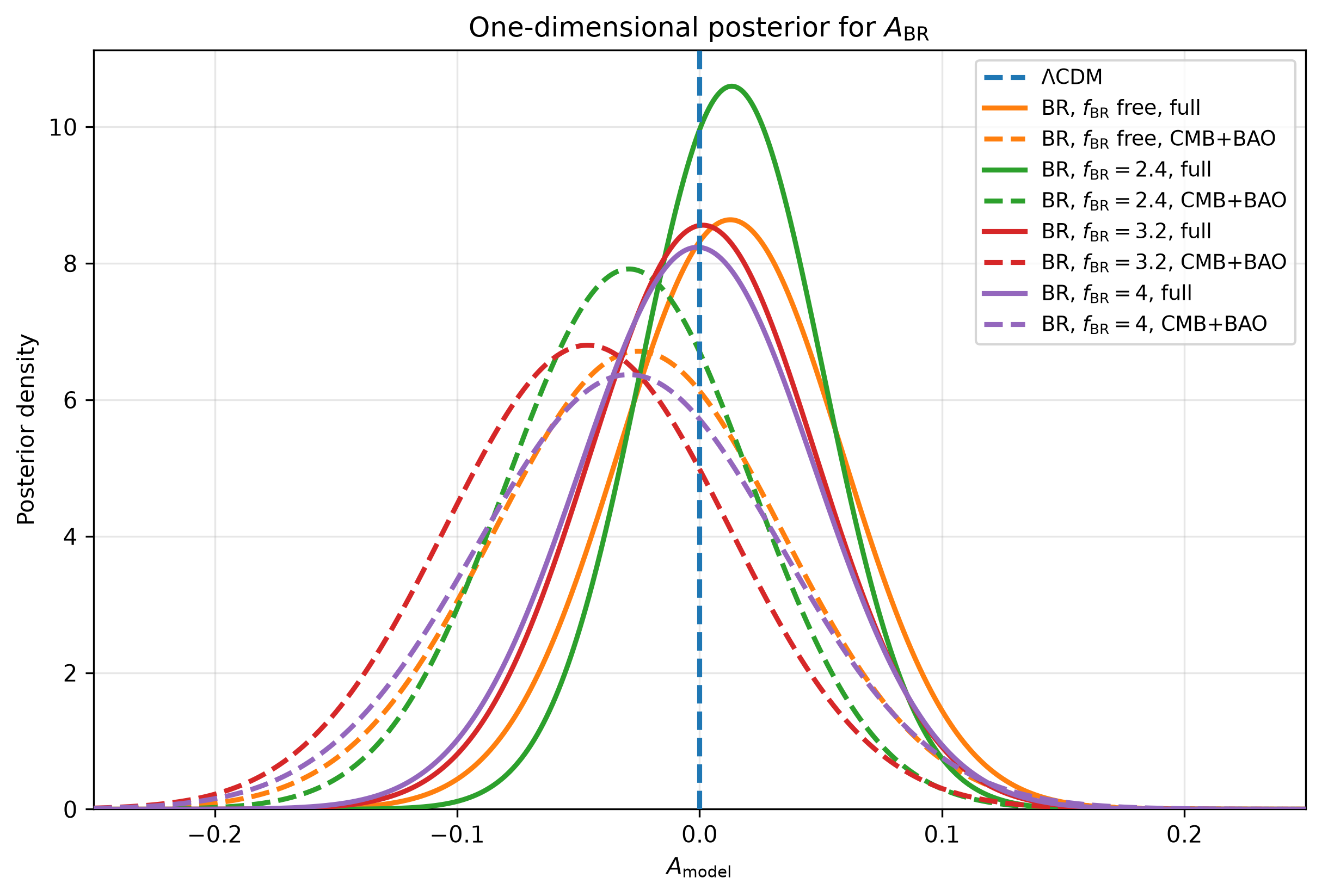}
\caption{Posterior probability for the parameter $A_{\rm BR}$ using the full set of data (solid curves), and only the CMB and BAO data (dotted curves).}
\label{fig5}
\end{figure}

\section{Results and Discussion}
\label{sec:Results}

The constraints on the cosmological and model parameters are shown in figure~\ref{fig4} and listed in table~\ref{tablefits}. Our first key observation is that values of the constant $A_{\rm BR}$ up to a value $|A_{\rm BR}| < 0.1$ are consistent with the data.

Looking in more detail, we find that the parameter $A_{\rm BR}$ is constrained both when $f_{\rm BR}$ is fixed and when it is allowed to vary. In all cases, both for the full data set and for the CMB+BAO combination, the inferred value of $A_{\rm BR}$ is compatible with zero. From figure~\ref{fig5}, we see that the full data set mildly prefers positive values of $A_{\rm BR}$, whereas the CMB+BAO data alone favour slightly negative values. Moreover, figure~\ref{fig4} shows a weak correlation between $A_{\rm BR}$ and the present-day value of the Hubble parameter, $H_0$. Finally, the inclusion of the full data set reduces the bound on $A_{\rm BR}$.

Allowing $f_{\rm BR}$ to vary does not significantly affect the constraint on $A_{\rm BR}$. However, the data are not able to constrain $f_{\rm BR}$ itself, which remains essentially unconstrained. The remaining cosmological parameters are not significantly shifted by the inclusion of the back-reaction mechanism.

Although $f_{\rm BR}$ is not directly constrained by the posterior distribution, we can still assess the relative statistical preference for the different models through the Deviance Information Criterion (DIC), reported in the last column of table~\ref{tablefits}. 
The DIC criteria \cite{Spiegelhalter2002} is commonly used for Bayesian model comparison from MCMC samples and has also been adopted in cosmological applications \cite{Liddle2007}. The DIC of the vector of free parameters $\theta$ is defined as
\begin{equation}
    {\rm DIC} = 2\langle \chi^2 \rangle - \chi^2(\bar{\theta}) ,
\end{equation}
where $\langle \chi^2 \rangle$ is the posterior mean of the effective chi-square, and $\chi^2(\bar{\theta})$ is the effective chi-square evaluated at the posterior mean of the parameter vector $\bar{\theta}$. We define
\begin{equation}
    \Delta{\rm DIC} =
    {\rm DIC}_{\rm model} - {\rm DIC}_{\Lambda{\rm CDM}} ,
\end{equation}
so that negative values indicate a preference for the extended model with respect to $\Lambda$CDM.

Since the Jeffreys scale is formally defined for Bayes factors rather than for information criteria, we adopt here a Jeffreys-like empirical interpretation of DIC differences: $|\Delta{\rm DIC}|<1$ is regarded as statistically negligible, $2\leq |\Delta{\rm DIC}|<5$ as positive evidence, and $|\Delta{\rm DIC}|\geq 5$ as strong evidence in favour of the model with the smaller DIC.

For the full data set, the model in which both $A_{\rm BR}$ and $f_{\rm BR}$ are allowed to vary gives $\Delta{\rm DIC}=-2.5$. Since lower values of the DIC correspond to a better balance between goodness of fit and model complexity, this negative value indicates a preference for the back-reaction model over $\Lambda$CDM.
According to the empirical criterion adopted above, this corresponds to positive evidence in favour of the back-reaction model.

On the other hand, the back-reaction model with fixed $f_{\rm BR}=3.2$ gives $\Delta{\rm DIC}=1.7$. With our convention, this positive value favours $\Lambda$CDM over the fixed-$f_{\rm BR}$ back-reaction model. However, since the difference is smaller than the threshold for positive evidence, this preference should be regarded as statistically weak. This result is nevertheless
interesting because, as shown in figure~\ref{fig3}, the parameter $f_{\rm BR}$ affects the ISW contribution.

\begin{table}[htbp]
\centering
\setlength{\tabcolsep}{4pt}
\renewcommand{\arraystretch}{1.5}
\resizebox{\textwidth}{!}{%
\begin{tabular}{ccccccccc}
\hline\hline
Model 
& Dataset 
& $A_{\rm BR}$ 
& $f_{\rm BR}$ 
& $\Omega_{m}$ 
& $10^9A_s$ 
& $H_0 \, (\mathrm{km \, s^{-1} \, Mpc^{-1}})$ 
& $\sigma_8$ 
& $\Delta{\rm DIC}$ \\
\hline

$\Lambda \mathrm{CDM}$ 
& full 
& - 
& - 
& $0.3064 \pm 0.0053$ 
& $2.111 \pm 0.031$ 
& $68.01 \pm 0.40$ 
& $0.8088 \pm 0.0060$ 
& 0 \\

$\Lambda \mathrm{CDM}$ 
& CMB+BAO 
& - 
& - 
& $0.3108 \pm 0.0057$ 
& $2.104 \pm 0.029$ 
& $67.68 \pm 0.43$ 
& $0.8094 \pm 0.0058$ 
& 0 \\

BR 
& full 
& $0.013 \pm 0.038$ 
& $2.4$ 
& $0.3069 \pm 0.0054$ 
& $2.111 \pm 0.031$ 
& $68.09 \pm 0.46$ 
& $0.8103 \pm 0.0070$ 
& $1.4$ \\

BR 
& CMB+BAO 
& $-0.029 \pm 0.050$ 
& $2.4$ 
& $0.3108 \pm 0.0059$ 
& $2.105 \pm 0.030$ 
& $67.44 \pm 0.61$ 
& $0.8075 \pm 0.0074$ 
& $1.4$ \\

BR 
& full 
& $0.001 \pm 0.047$ 
& $3.2$ 
& $0.3064 \pm 0.0055$ 
& $2.110 \pm 0.031$ 
& $68.03 \pm 0.46$ 
& $0.8087 \pm 0.0065$ 
& $1.7$ \\

BR 
& CMB+BAO 
& $-0.046 \pm 0.059$ 
& $3.2$ 
& $0.3113 \pm 0.0059$ 
& $2.103 \pm 0.030$ 
& $67.41 \pm 0.56$ 
& $0.8074 \pm 0.0067$ 
& $1.3$ \\

BR 
& full 
& $-0.0009 \pm 0.0484$ 
& $4$ 
& $0.3065 \pm 0.0054$ 
& $2.109 \pm 0.030$ 
& $68.01 \pm 0.44$ 
& $0.8086 \pm 0.0062$ 
& $1.5$\\

BR 
& CMB+BAO 
& $-0.029 \pm 0.063$ 
& $4$ 
& $0.3113 \pm 0.0057$ 
& $2.101 \pm 0.028$ 
& $67.54 \pm 0.51$ 
& $0.8083 \pm 0.0061$ 
& $0.4$ \\

BR, $f_{\rm BR}$ free  
& full 
& $0.013 \pm 0.046$ 
& unconstrained 
& $0.3074 \pm 0.0056$ 
& $2.107 \pm 0.030$ 
& $68.12 \pm 0.49$ 
& $0.8103 \pm 0.0068$ 
& $-2.5$ \\

BR, $f_{\rm BR}$ free  
& CMB+BAO 
& $-0.025 \pm 0.059$ 
& unconstrained 
& $0.3111 \pm 0.0059$ 
& $2.103 \pm 0.028$ 
& $67.55 \pm 0.60$ 
& $0.8084 \pm 0.0070$ 
& $0.7$ \\

\hline
\end{tabular}%
}
\caption{Marginalized posterior means and 1\,$\sigma$ standard deviations for selected cosmological parameters and derived quantities of the models considered in this work. The last column reports the $\Delta{\rm DIC}$, calculated respect to the $\Lambda$CDM model fitted to the same dataset, using ${\rm DIC}=2\langle\chi^2\rangle-\chi^2(\bar{\theta})$.}
\label{tablefits}
\end{table}

\section{Conclusion}

In this work we have investigated a phenomenological back-reaction-motivated interacting dark-sector scenario in which the effective energy exchange between the matter sector and the back-reaction contribution oscillates as a function of redshift. The model modifies the late-time background evolution while matching the standard matter behaviour before recombination.  As a consequence, the main CMB signature is expected at low multipoles, through a modification of the late ISW contribution. 

We have found that the data allows for a significant contribution of the back-reaction terms: values of $|A_{\rm BR}| \leq 0.1$ are allowed at the one sigma level. For these values of $A_{\rm BR}$, the effect of the back-reaction term of the value of today's Hubble constant is small and does not effect the Hubble tension issue.

Using CMB, BAO and supernova data, we find that the amplitude parameter $A_{\rm BR}$ is consistent with zero in all the cases considered. The inclusion of the full data set reduces the uncertainty on $A_{\rm BR}$ and mildly shifts its preferred value towards positive values, whereas the CMB+BAO combination alone mildly prefers negative values.  One must have in mind that the signal of $A_{\rm BR}$ indicates the signal  of the effective back-reaction  density for $z \sim 0$.  Note that in this work the parameter $A_{\rm BR}$ is taken to be a constant, independent of redshift, and as the redshift  increases the physical amplitude of the effective back-reaction density  oscillates  between positive and negative values.  The remaining cosmological parameters, including $H_0$, $\Omega_m$ and $\sigma_8$, remain statistically consistent with their corresponding $\Lambda$CDM values. When $f_{\rm BR}$ is allowed to vary, the data do not provide a direct posterior constraint on this parameter. Nevertheless, the DIC comparison shows that the full data set has a preference of the back-reaction model. By contrast, the models with fixed $f_{\rm BR}=3.2$ has $\Delta{\rm DIC}=1.7$, which mildly favours the $\Lambda$CDM model, but remains below the threshold for positive evidence. These results suggest that the data are not primarily selecting a fixed oscillation frequency, but rather allow for a weak preference for a broader class of oscillatory back-reaction histories when the frequency is left free. This is interesting because the parameter $f_{\rm BR}$ controls the redshift structure of the oscillations and, consequently, the angular scale at which the low-$\ell$ CMB temperature spectrum is affected. 

A more complete treatment, including a detailed analysis of the evolution of the perturbations of the  back-reaction component and a dedicated analysis of the ISW contribution, will be necessary to determine whether this mild preference can be associated with a robust physical signature. Given that $\sigma_8$ and the growth rate can respond to a time-oscillating equation of state even when the background expansion is only weakly perturbed, extending the perturbative sector to include the back-reaction fluid is a natural continuation of this work, and one we consider necessary before drawing firm conclusions on the viability of the scenario.

The oscillatory form assumed in eq.~\eqref{eq:rhoBR_def_ref} was motivated qualitatively by the back-reaction mechanism discussed in appendix~\ref{App:A}, but a quantitative bridge between the two is not yet in place: in particular, a derivation of the values of $A_{\rm BR}$ and $f_{\rm BR}$ expected from a given spectrum of super-Hubble fluctuations generated during inflation would let the constraints obtained here bear directly on the underlying mechanism, rather than on its phenomenological parametrization. We leave this connection for future work. An interesting extension of our work would be to consider a possible redshift dependence of the amplitude $A_{\rm BR}$.

\acknowledgments
 
We wish to thank Elisa Ferreira for extensive discussions and E. Colgain for important feedback on the first draft of this paper.  M.A.C.A. is supported by Coordenacao de Aperfeicoamento de Pessoal de Nivel Superior (CAPES). M.A.C.A. would like to thank Scuola Superiore Meridionale (Naples) for warm hospitality during the period that part of this research was developed. M.B. acknowledges Istituto Nazionale di Fisica Nucleare (INFN), sezione di Napoli, \textit{iniziative specifica} QGSKY.
L.L.G is supported by research grants from Conselho Nacional de Desenvolvimento Cientifico e Tecnologico (CNPq),  Grant No. 307636/2023-2 and from the Fundacao Carlos Chagas Filho de Amparo a Pesquisa do Estado do Rio de Janeiro (FAPERJ), Grant No. E-26/204.598/2024. L.L.G. would like to thank Scuola Superiore Meridionale (Naples) for warm hospitality during the period that part of this research was developed.  R.B. is supported in part by NSERC and by funds from McGill University.

\appendix
\section{Back-reaction motivation for an oscillating dark sector}\label{App:A}

In this appendix we review the motivation for an oscillating effective equation of state for the joint dark matter - dark energy fluid which comes from back-reaction considerations \cite{Brandenberger:2002sk}. The starting point of the scenario was the realization \cite{Mukhanov:1996ak,Abramo:1997hu} that long wavelength (super Hubble) cosmological perturbations act as a negative contribution to the effective cosmological constant. The physical reason for this is easy to understand: Since the Einstein equations are nonlinear, cosmological perturbations lead to back-reaction contributions to the background, at quadratic order in the amplitude of the fluctuations.  Importantly, matter fluctuations induce gravitational potential wells.  On super-Hubble scales the negativity of the gravitational contribution to the effective energy overcomes the positive contribution from the matter fluctuation. Gradient contributions to the energy are negligible on super-Hubble scales, and kinetic contributions are suppressed because the Newtonian gravitational potential is conserved on super-Hubble scales (see \cite{MFB, RHBfluctsrev} for reviews on the theory of cosmological perturbations). Hence, the equation of state of the induced effective energy-momentum tensor by which the fluctuations back-react on the metric is that of a negative cosmological constant.

Unruh raised a key question \cite{Unruh} as to whether the above back-reaction effect is physically measurable. In fact, in models in which the clock field which we use to interpret observations is set by the dominant fluid, the back-reaction effect is not measurable \cite{GeshnizjaniBrandenberger2002,AbramoWoodard2002NoOneLoop,AfshordiBrandenberger2001}. However,  if the dominant matter fluid fluctuates on the hypersurface of constant clock field, as it does in our late universe, then the back-reaction effect is measurable \cite{Geshnizjani:2003cn, Brandenberger:2018fdd, Comeau:2023euf, AbramoWoodard2002BackReaction,LosicUnruh2005,LosicUnruh2006,Marozzi:2012ib,Brandenberger:2022hmf}.

If one considers a model with a bare cosmological constant, it will generate a period of accelerated expansion during which fluctuation modes exit the Hubble radius. The phase space of long wavelength modes grows and the induced negative contribution to the effective cosmological constant $\lambda_{eff}$ grows in proportion. Once the magnitude of the energy density associated with $\Lambda_{eff}$ becomes smaller than the matter energy density (or whatever form of matter dominates the Universe) $\rho_m(t)$, the  accelerated expansion will stop, modes will re-enter the Hubble radius and hence the magnitude of $\Lambda_{eff}$ will stop decreasing. The timescale for this dynamics is the Hubble timescale. Since $\Lambda_{eff}$ no longer decreases while  $\rho_m(t)$ is decreasing, a new phase of acceleration will start, and then $\Lambda_{eff}$ will commence to decrease again. Thus, on a Hubble timescale, intervals of dark energy domination and matter domination will follow each other.  

Translated to the general scenario we describe in the main text, there is an exchange of energy between the effective cosmological constant sector and the matter sector which oscillates on a Hubble time scale.  Note that we do not have an initial accelerating phase in which the large phase space of super-Hubble fluctuation modes builds up.  If we are to connect our oscillating dark sector model to the back-reaction scenario described above, we would have to assume that there is already a large phase space of super-Hubble modes present (e.g. built up in a phase of inflation in the very early universe).  In addition, for the small amplitude of $A_{\rm BR}$ considered in this paper, we would never have accelerating phases.  Nevertheless, the period of oscillation might be a remnant of oscillations of the type described in this appendix which occurred in the early universe.  We leave the establishment of a more direct link between the back-reaction scenario described here and the parametrization we use in the main text for future work.

Note: we wish to draw attention to a couple of recent papers \cite{Comeau:2026spq, Macpherson:2026bzj,  Koksbang:2026tfw} concerning back-reaction and interacting dark sector models.

\section{Exact background solution}\label{app:B}

The integrating factor for equation~\eqref{eq:y_ODE_m_case} is
\begin{equation}\label{eq:mu_integrating_factor}
\mu(z)=\exp\!\left[-\int\frac{3}{1+z}\,dz\right]=(1+z)^{-3}.
\end{equation}
Thus,
\begin{equation}\label{eq:y_integrating_factor_step}
\begin{split}
    \dv{}{z}\!\left[(1+z)^{-3}y(z)\right]&=-A_{\rm BR}\bigg[\frac{3\cos(f_{\rm BR}z)}{1+z}
    -f_{\rm BR}\sin(f_{\rm BR}z)\bigg].
\end{split}
\end{equation}
Integrating from $z_{\rm rec}=1089.92$ to $z$ yields
\begin{equation}\label{eq:y_general_solution_before_matching}
\begin{split}
y(z)&=(1+z)^3\!\bigg[\frac{y_{\rm rec}}{(1+z_{\rm rec})^3}+A_{\rm BR}\big(\cos(f_{\rm BR}z_{\rm rec})-\cos(f_{\rm BR}z)\big)\\
&-3A_{\rm BR}\!\int_{z_{\rm rec}}^{z}\frac{\cos(f_{\rm BR}z')}{1+z'}\,dz'
\bigg].
\end{split}
\end{equation}
We then impose that, at recombination, the total matter-like sector matches the reference $\Lambda$CDM matter density:
\begin{equation}\label{eq:recombination_matching_condition}
\rho_{\rm M}(z_{\rm rec})
=\rho_{\rm m0}(1+z_{\rm rec})^3.
\end{equation}
Using $\rho_{\rm M}=\rho_{\rm m}+\rho_{\rm BR}$ and eqs.~\eqref{eq:rhoBR_def_ref} and \eqref{eq:y_general_solution_before_matching}, this becomes
\begin{equation}\label{eq:yrec_matching_equation}
y_{\rm rec}+A_{\rm BR}\cos(f_{\rm BR}z_{\rm rec})(1+z_{\rm rec})^3=(1+z_{\rm rec})^3,
\end{equation}
hence
\begin{equation}\label{eq:yrec_solution}
\boxed{\,y_{\rm rec}=(1+z_{\rm rec})^3\Big[1-A_{\rm BR}\cos(f_{\rm BR}z_{\rm rec})\Big].\,}
\end{equation}
Substituting eq.~\eqref{eq:yrec_solution} back into eq.~\eqref{eq:y_general_solution_before_matching} yields
\begin{gather}\label{eq:y_in_terms_of_I}
\boxed{
y(z)=(1+z)^3\left[1-A_{\rm BR}\cos(f_{\rm BR}z)-3A_{\rm BR}\,I(z)\right]},
\\
I(z)\equiv\int_{z_{\rm rec}}^{z}\frac{\cos(f_{\rm BR}z')}{1+z'}\,dz'.
\end{gather}
Now let $u=1+z'$ so that $dz'=du$ and
$\cos(f_{\rm BR}z')=\cos[f_{\rm BR}(u-1)]=\cos(f_{\rm BR}u)\cos (f_{\rm BR})+\sin(f_{\rm BR}u)\sin (f_{\rm BR})$.
Then, with $t=f_{\rm BR}u$,
\begin{equation}\label{eq:I_u_change_of_vars}
\begin{split}
    I(z)&=\cos (f_{\rm BR})\int_{f_{\rm BR}(1+z_{\rm rec})}^{f_{\rm BR}(1+z)}\frac{\cos t}{t}\,dt+\sin (f_{\rm BR})\int_{f_{\rm BR}(1+z_{\rm rec})}^{f_{\rm BR}(1+z)}\frac{\sin t}{t}\,dt.
\end{split}
\end{equation}
Using the standard definitions $\Si'(x)=\sin x/x$ and $\Ci'(x)=\cos x/x$, we obtain
\begin{equation}\label{eq:I_exact}
\boxed{
\begin{aligned}
    I(z)=&\cos (f_{\rm BR})\Big[\Ci\!\big(f_{\rm BR}(1+z)\big)-\Ci\!\big(f_{\rm BR}(1+z_{\rm rec})\big)\Big]\\
    &+\sin (f_{\rm BR})\Big[\Si\!\big(f_{\rm BR}(1+z)\big)-\Si\!\big(f_{\rm BR}(1+z_{\rm rec})\big)\Big].
\end{aligned}
}
\end{equation}

Therefore, the exact solution is
\begin{gather}\label{eq:y_exact_app}
\boxed{
\begin{aligned}
    y(z)&=(1+z)^3\!\{
    1-A_{\rm BR}\cos(f_{\rm BR}z)\\
    &-3A_{\rm BR}[\cos (f_{\rm BR})\,\left(\Ci(f_{\rm BR}(1+z))-\Ci(f_{\rm BR}(1+z_{\rm rec}))\right)\\
    &+\sin (f_{\rm BR})\,\left(\Si(f_{\rm BR}(1+z))-\Si(f_{\rm BR}(1+z_{\rm rec}))\right)
]
\}
\end{aligned}
}
\end{gather}

\bibliographystyle{JHEP}
\bibliography{bibliography_draft_JCAP}

\end{document}